\documentclass[PAPER, cernpreprint, atlasdraft=false, UKenglish, texmf, orcidlogo]{atlasdoc}
\usepackage{atlaspackage}
\usepackage{atlasbiblatex}

\usepackage{atlasphysics}
\graphicspath{{logos/}{figures/}}

\usepackage{ANA-TOPQ-2023-03-PAPER-defs}

\AtlasTitle{Observation of $t\bar{t}\gamma\gamma$ production at $\sqrt{s}=$13 TeV with the ATLAS detector}

\AtlasAbstract{%
This paper presents the first observation of top-quark pair production in association with two photons ($t\bar{t}\gamma\gamma$). The measurement is performed in the single-lepton decay channel using proton–proton collision data collected by the ATLAS detector at the Large Hadron Collider. The data correspond to an integrated luminosity of 140\,fb$^{-1}$ recorded during Run 2 at a centre-of-mass energy of 13 TeV. The $t\bar{t}\gamma\gamma$ production cross section, measured in a fiducial phase space based on particle-level kinematic criteria for the lepton, photons, and jets, is found to be $2.42^{+0.58}_{-0.53}\, \text{fb}$, corresponding to an observed significance of 5.2 standard deviations. Additionally, the ratio of the production cross section of $t\bar{t}\gamma\gamma$ to top-quark pair production in association with one photon is determined, yielding $(3.30^{+0.70}_{-0.65})\times 10^{-3}$.
}

\AtlasRefCode{TOPQ-2023-03}

\PreprintIdNumber{CERN-EP-2025-125}

\AtlasJournalRef{\PLB 874 (2026) 140195}
\AtlasDOI{10.1016/j.physletb.2026.140195}

\AtlasCoverEgroupAnalysisTeam{atlas--analysis-team@cern.ch}

\hypersetup{pdftitle={ATLAS document},pdfauthor={The ATLAS Collaboration}}

\begin{document}

\maketitle

\section{Introduction}

Measurements of top-quark pair production (\ttbar) in association with neutral vector bosons ($\gamma$, $Z$ boson) allow a direct probe of the electroweak couplings of the top quark. In particular, the associated production of top quarks with photons is a probe of the top-quark--photon coupling. These measurements provide access to the electric
charge and the electroweak dipole moments of the top quark and, more generally, allow constraints on modifications of the structure of these couplings relative to the Standard Model predictions~\cite{Baur:2004uw, Bouzas:2012av, Schulze:2016qas,Bylund:2016phk}.

The production of \ttbar in association with a single photon (\tty), for which evidence was first reported by the CDF Collaboration~\cite{Aaltonen:2011sp}, has been studied in detail by the ATLAS and CMS Collaborations. Inclusive and differential production cross-section measurements were performed at centre-of-mass energies ($\sqrt{s}$) ranging from 7 to 13~\TeV, in both the dilepton and single-lepton \ttbar decay channels~\cite{TOPQ-2012-07, TOPQ-2015-21, CMS-TOP-14-008, TOPQ-2017-14, TOPQ-2020-03, CMS-TOP-18-010, CMS-TOP-21-004}. Recent measurements by the ATLAS Collaboration focus specifically on the production of \tty events, where the photon arises from initial-state radiation or an off-shell top quark, measuring inclusive and differential cross sections~\cite{TOPQ-2018-24} and the \ttbar charge asymmetry in \tty events~\cite{TOPQ-2021-29}. However, the associated production of \ttbar events with two photons (\ttyy), a much rarer process with an expected production cross section at the per mille level compared to \tty, was not specifically studied. In \ttyy final states, the photons can be emitted by any charged particle: the initial-state partons, the top quarks or the top quark decay products (including the $W$ boson and $W$ boson decay products). Example Feynman diagrams for \ttyy production are shown in Figure~\ref{fig:Feynman}. The \ttyy process has been discussed in several recent calculations with stable top quarks and including the decays at parton level~\cite{Pagani:2021,Stremmer:2023kcd,Stremmer:2024ecl}. It represents a relevant irreducible background for the measurements of the production of the Higgs boson in association with \ttbar (\ttH), where the Higgs boson decays into two photons. It could additionally be a handle to further constrain the anomalous chromomagnetic and chromoelectric dipole moments~\cite{PhysRevD.100.015023}. The ratio of the cross section between the \ttyy and \tty processes is also proposed as an observable to constrain the anomalous dipole moments.

This \paperornote presents the first measurement of the \ttyy fiducial cross section and the ratio of the \ttyy and \tty cross sections. Processes such as \tty and \ttyy, including photons radiated from any charged particles in the initial
or final states, cannot currently be simulated at high-order accuracy using matrix-element (ME) generators interfaced with parton shower (PS) simulations. %
Moreover, the previously mentioned \ttyy theoretical calculations cannot be directly compared with measurements. Therefore, the main goal of this \paperornote is to provide a first reference measurement for future theoretical developments, instead of putting the emphasis on comparisons with the existing simulations. The measurements are performed by selecting \ttbar single-lepton final states, characterised by one high-transverse-momentum (\pt) lepton and at least four jets, two of which arise from $b$-quarks ($b$-jets), and by requiring the presence of exactly two high-\pt photons. They are measured using the full data set recorded by the ATLAS detector at the Large Hadron Collider (LHC) between 2015 and 2018 at $\sqrt{s}=13~\TeV$, corresponding to an integrated luminosity of 140~\ifb~\cite{DAPR-2021-01}. The background processes comprise events with genuine photons and events with fake photons. These processes are estimated by using a combination of simulation and data-driven methods. The output of a boosted decision tree (BDT) classifier, trained to separate signal and background processes, is used as the observable for the measurements. In particular, the cross section is measured in a fiducial phase space at particle level using a profile likelihood fit to the BDT output. The fiducial phase space is defined by requiring exactly two photons, one electron or muon, and at least four jets, including at least one $b$-jet, closely following the kinematic requirements at reconstruction level. Additionally, the ratio of the \ttyy to \tty cross sections is extracted from a simultaneous measurement, incorporating data and simulation inputs from the \tty analysis described in Ref.~\cite{TOPQ-2018-24}. This measurement is expected to yield better precision than the cross section owing to the cancellation of normalisation uncertainties.

\begin{figure}
\centering
\subfloat[]{\includegraphics[width=0.45\textwidth]{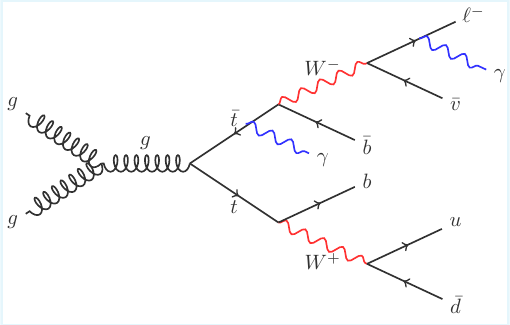}}%
\subfloat[]{\includegraphics[width=0.45\textwidth]{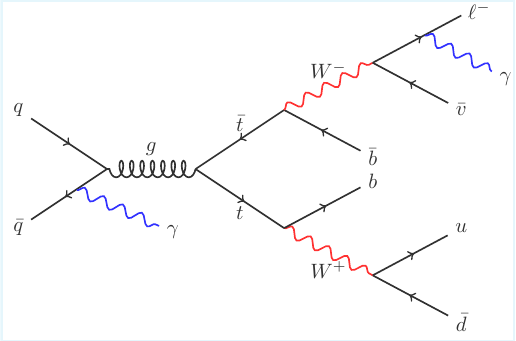}}
\caption{Examples of leading-order Feynman diagrams for \ttyy production in the single-lepton \ttbar final state, where the photons are radiated by (a) an off-shell top quark and the charged lepton and (b) an initial-state quark and the charged lepton, respectively. }
\label{fig:Feynman}
\end{figure}
\section{ATLAS detector}
\newcommand{\AtlasCoordFootnote}{%
ATLAS uses a right-handed coordinate system with its origin at the nominal interaction point (IP)
in the centre of the detector and the \(z\)-axis along the beam pipe.
The \(x\)-axis points from the IP to the centre of the LHC ring,
and the \(y\)-axis points upwards.
Polar coordinates \((r,\phi)\) are used in the transverse plane,
\(\phi\) being the azimuthal angle around the \(z\)-axis.
The pseudorapidity is defined in terms of the polar angle \(\theta\) as \(\eta = -\ln \tan(\theta/2)\) and is equal to the rapidity
$ y = \frac{1}{2} \ln \left( \frac{E + p_z}{E - p_z} \right) $ in the relativistic limit.
Angular distance is measured in units of \(\Delta R \equiv \sqrt{(\Delta y)^{2} + (\Delta\phi)^{2}}\).}

The ATLAS experiment~\cite{PERF-2007-01} at the LHC is a multipurpose particle detector with a forward--backward symmetric cylindrical geometry and a near \(4\pi\) coverage in
solid angle.\footnote{\AtlasCoordFootnote}
It consists of an inner tracking detector (ID) surrounded by a thin superconducting solenoid
providing a \qty{2}{\tesla} axial magnetic field, electromagnetic and hadronic calorimeters, and a muon spectrometer.
The inner tracking detector covers the pseudorapidity range \(|\eta| < 2.5\).
It consists of silicon pixel, silicon microstrip, and transition radiation tracking detectors.
Lead/liquid-argon (LAr) sampling calorimeters provide electromagnetic (EM) energy measurements
with high granularity within the region \(|\eta|< 3.2\).
A steel/scintillator-tile hadronic calorimeter covers the central pseudorapidity range (\(|\eta| < 1.7\)).
The endcap and forward regions are instrumented with LAr calorimeters
for EM and hadronic energy measurements up to \(|\eta| = 4.9\).
The muon spectrometer surrounds the calorimeters and is based on
three large superconducting air-core toroidal magnets with eight coils each.
The field integral of the toroids ranges between \num{2.0} and \qty{6.0}{\tesla\metre}
across most of the detector.
The muon spectrometer includes a system of precision tracking chambers up to \(|\eta| = 2.7\) and fast detectors for triggering up to \(|\eta| = 2.4\).
The luminosity is measured mainly by the LUCID--2~\cite{LUCID2} detector, which is located close to the beampipe.
A two-level trigger system is used to select events~\cite{TRIG-2016-01}.
The first-level trigger is implemented in hardware and uses a subset of the detector information
to accept events at a rate below \qty{100}{\kHz}.
This is followed by a software-based trigger that
reduced the accepted rate of complete events to \qty{1.25}{\kHz} on average in Run~2 depending on the data-taking conditions. A software suite~\cite{SOFT-2022-02} is used in data simulation, in the reconstruction
and analysis of real and simulated data, in detector operations, and in the trigger and data acquisition
systems of the experiment.

\section{Modelling of signal and background processes}
\label{sec:simulations}
The signal and background processes are modelled using Monte Carlo (MC) simulation samples. The ATLAS detector response is simulated~\cite{SOFT-2010-01} with \GEANT~\cite{Agostinelli:2002hh}, and some alternative samples for evaluating systematic uncertainties are processed with a fast simulation (\textsc{AtlFast-II}) relying on calorimeter response parameterisation. All the MC samples  described in the following that were generated with either \powhegbox~\cite{Nason:2004rx,Frixione:2007nw,Frixione:2007vw,Alioli:2010xd} or \MGNLO~\cite{Alwall:2014hca} were interfaced to \PYTHIA[8]~\cite{Sjostrand:2014zea} to simulate the PS, fragmentation, and underlying event. The A14 set of tuned parameters (tune)~\cite{ATL-PHYS-PUB-2014-021} and the \NNPDF[2.3lo] parton distribution function (PDF) set~\cite{Ball:2012cx} were used in \Pythia. The heavy-flavour hadron decays were modelled by EvtGen~\cite{Lange:2001uf}. Samples generated with \Sherpa~\cite{Gleisberg:2008ta,Hoeche:2009rj} used the \Sherpa parton shower with a dedicated tune provided by the authors~\cite{Schumann:2007mg}.
The top quark mass was set to \SI{172.5}{\GeV} in the samples. Pile-up effects, additional \pp~collisions in the same or neighbouring bunch crossings, were simulated by overlaying minimum bias events using \PYTHIA[8]~\cite{Sjostrand:2007gs} with the A3 tune~\cite{ATL-PHYS-PUB-2016-017}. The MC events were reweighted to match the observed number of interactions per bunch crossing. Additionally, corrections for trigger, reconstruction and selection efficiencies, and energy scales are applied to MC events to improve the description of the data.

The signal \ttyy process was simulated as a $2\to 8$ process\footnote{The 8 final-state particles correspond to the 6 decay products of the \ttbar system and the 2 photons.} with \MGNLO[2.9.9] at leading-order (LO) in quantum chromodynamics (QCD) with the \NNPDF[3.0nlo] set of PDFs~\cite{Ball:2014uwa}. Diagrams where photons are radiated from the initial state, the top quarks, or their charged decay products are considered. The event generation is interfaced to \PYTHIA[8.306]~\cite{Bierlich:2022pfr}. The renormalisation and factorisation scales were set to $0.5\times \sum_i \sqrt{m^2_i+p^2_{\text{T},i}}$, where $m_i$ and $p_{\text{T},i}$ are the masses and transverse momenta of the particles generated from the ME calculation. Photons were required to have $\pT > 15~\GeV$ and to be isolated according to a smooth-cone hadronic isolation criterion with $\delta_0=0.1$, $\epsilon_{\gamma}=0.1$ and $n=2$, defined in Ref.~\cite{Frixione:1998jh}, to avoid infrared divergences. Leptons and quarks at the ME level are required to have a minimum transverse momentum of 4~\GeV and 1~\GeV respectively. The normalisation of the LO $2\to 8$ \ttyy sample is rescaled using a $K$-factor calculated from two $pp \rightarrow$ \ttyy samples ($2\rightarrow 4$), in which photons are emitted from either off-shell top quarks or initial-state particles. This process can also be simulated at next-to-leading-order (NLO) in QCD with \MGNLO, unlike the $2\rightarrow 8$ \ttyy sample. The NLO and LO $2\rightarrow 4$ \ttyy samples are simulated using \MGNLO interfaced with \pythia8, with the same parameters as the nominal sample. The expected cross sections in the fiducial phase space at particle level described in Section~\ref{sec:measurements} are $0.733^{+0.073}_{-0.080}(\text{scale})^{+0.016}_{-0.008}(\text{PDF})\,\text{fb}$ and $0.423^{+0.070}_{-0.090}(\text{scale})^{+0.009}_{-0.004}(\text{PDF})\,\text{fb}$ for the NLO and LO $2\to 4$ simulations, respectively, corresponding to a $K$-factor of 1.7.

Top-quark-related processes constitute a background to the \ttyy signal, as their final states result in topologies with a lepton, jets, $b$-jets, and either genuine photons or fake photons from misreconstructed objects. The production of vector bosons in association with photons constitutes a background as well, as these events may include $b$-jets from initial-state radiation. These processes are simulated as described below.

The \tty events were simulated as two complementary samples using also the \MGNLO generator. The first, referred to as `\tty production sample’, models photon production from the off-shell top quark or initial-state radiation in a $2 \to 3$ process at NLO in QCD, with top quark decays at LO using \MADSPIN~\cite{Frixione:2007zp,Artoisenet:2012st}. The second, the `\tty decay sample’, simulates photon production from on-shell top quarks and their decay products as a $2 \to 2$ process at LO\@. Both samples use the \NNPDF[3.0nlo] PDF set. The same photon requirements and renormalisation and factorisation scales as in the \ttyy sample are set in these samples. The \tty production sample is normalised to the NLO cross section given by the MC simulation, while the normalisation of the \tty decay sample is corrected by an NLO/LO inclusive $K$-factor of 1.5, as obtained in Ref.~\cite{TOPQ-2021-29}.

The production of \ttbar and single-top-quark events ($t$- and $s$-channels) was modelled at NLO in QCD using \powhegbox~\cite{Nason:2004rx,Frixione:2007nw,Frixione:2007vw,Alioli:2010xd}. For the \ttbar sample, the \hdamp parameter, which controls the \pt of the first additional emission, was set to 1.5 times the top quark mass~\cite{ATL-PHYS-PUB-2016-020}. The \ttbar and single-top-quark simulation samples are normalised to the cross sections calculated at next-to-next-to-leading-order (NNLO) in QCD including the
resummation of next-to-next-to-leading-logarithmic (NNLL) soft-gluon terms~\cite{Czakon:2011xx} at NNLO or approximated NNLO~\cite{Kidonakis:2010ux,Kidonakis:2010tc,Kidonakis:2011wy}, respectively. The \tWy events were generated at LO with \MGNLO in the five-flavour scheme as two samples: one as a $2 \to 3$ process with a stable top quark and the other as a $2 \to 2$ process with photon radiation from final-state charged particles. To avoid infrared divergences, the photon was required to have $\pT > 15~\GeV$ and $|\eta| < 5.0$ and to be separated by $\Delta R>0.2$ from any parton. Both \tWy samples use the \NNPDF[2.3lo] PDF set and are normalised to MC-predicted cross sections. Events with a \ttbar pair produced in association with a $W$~or $Z$~boson ($t\bar{t}V$) were simulated at NLO in QCD with \MGNLO using the \NNPDF[3.0nlo] PDF set. The simulation of \ttH events, with the Higgs boson decaying into two photons, was also performed at NLO in QCD with \MGNLO using the \NNPDF[3.0nlo] PDF set and interfaced to \pythia8.230. The sample is normalised to the total \ttH NLO cross section times the branching fraction of Higgs boson decays into two photons~\cite{deFlorian:2016spz}. %

Events with $V$ ($Z$ and $W$) bosons produced in association with one or two photons simulated at ME level, $V\gamma$ and $V\gamma\gamma$, and the production of $V$+jets and diboson processes were simulated with different versions of \sherpa~\cite{Gleisberg:2008ta,Hoeche:2009rj} at NLO in QCD using the \NNPDF[3.0nnlo] PDF set. The simulations of $V\gamma\gamma$ and $V\gamma$ are normalised to the cross section provided by the MC simulations, the diboson processes are normalised to cross sections at NLO accuracy in QCD~\cite{Campbell:1999ah}, while the $V$ boson samples are normalised to cross sections at NNLO accuracy in QCD~\cite{STDM-2015-03}.

The analysis combines multiple MC samples where the photons are generated either at the ME or the PS step. In particular, the \ttyy and $V\gamma\gamma$ samples include two photons generated at the ME level with certain kinematic requirements on the photons. A second set of samples --- \tty production and decay, \Wty, and $V\gamma$ --- contain one photon simulated at the ME level and the second one at PS\@. Additionally, inclusive samples are used in which photons are generated only during the PS step. To avoid the double-counting of events among samples corresponding to the same physics process, the overlapping events in the samples without explicit photon inclusion at ME are removed. The overlap among events is considered for those with photons at parton level with $\pT > 15~\GeV$ and separated by $\Delta R>0.2$ from any charged lepton, referred to as the \textit{overlap region}. First, events from the inclusive samples without any photons at ME (\ttbar, $V$+jets) are vetoed if there is only one candidate photon which falls into the overlap region. No events from those samples are found in the overlap region with the samples with two photons at ME\@. Events from samples with one photon simulated at ME (\tty production and decay and $V\gamma$) are discarded if both photons fall into the overlap region with \ttyy and $V\gamma\gamma$. This approach removes the overlap among events with photons generated in the same phase space, while keeping the events from such samples that might satisfy the event selection at reconstruction level because a candidate object is a fake photon. For instance, an event from the \Vy samples with a genuine photon at ME in the overlap region and a fake photon at reconstruction level would not be removed.

\section{Event selection and background estimate}
The measurement is performed using the proton--proton collision data collected at the LHC between 2015 and 2018 at $\sqrt{s} = \SI{13}{\TeV}$ with the ATLAS detector, which corresponds to an integrated luminosity of 140~\ifb~\cite{DAPR-2021-01}. The event reconstruction, object identification and event selection follow closely that of the \tty cross-section measurement~\cite{TOPQ-2018-24}. In the following, a brief description of the event selection and background estimate is given. Further details can be found in Ref.~\cite{TOPQ-2018-24} and references therein.

Single-lepton triggers with electron and muon \pT thresholds between 20 and 27~\GeV depending on the data-taking period and various identification and isolation criteria~\cite{TRIG-2018-01,TRIG-2018-05,TRIG-2019-04,TRIG-2019-03} are used to select the events. Events are required to have at least one reconstructed collision vertex with two or more associated tracks with $\pt>0.5$~\GeV. The primary vertex is defined as the vertex with the largest $\sum \pt^2$ of the associated tracks~\cite{ATL-PHYS-PUB-2015-026} and consistent with the average beamspot position.

Photon candidates are reconstructed from energy deposits (clusters) in the electromagnetic calorimeter. Photons are required to satisfy a \emph{tight} identification criterion and fulfil calorimeter and track isolation criteria~\cite{EGAM-2018-01,TOPQ-2018-24} defined as $\left. E_{\mathrm{T}}^{\mathrm{iso}} \right|_{\Delta R = 0.4} < 0.022 \cdot E_{\mathrm{T}}(\gamma) + \SI{2.45}{\GeV}$ and $\left. \pT^{\mathrm{iso}} \right|_{\Delta R = 0.2} < 0.05 \cdot E_{\mathrm{T}}(\gamma)$. Here, $\left. E_{\mathrm{T}}^{\mathrm{iso}} \right|_{\Delta R = 0.4}$ corresponds to the calorimeter isolation within a cone of radius $\Delta R = 0.4$ in the direction of the photon candidate. The term $\left. \pT^{\mathrm{iso}} \right|_{\Delta R = 0.2}$ is the track isolation within $\Delta R = 0.2$ and $E_{\mathrm{T}}(\gamma)$ represents the transverse energy of a photon. Photon candidates are selected if they satisfy $E_{\mathrm{T}}(\gamma)>$~20~\GeV, $|\eta_{\text{cluster}}|<2.37$, excluding those in the transition region between the barrel and the endcap calorimeters (${|\eta_{\text{cluster}}|\not\in[1.37,1.52]}$). Photon candidates are classified as unconverted if the cluster is not matched to any reconstructed track in the ID system, or as converted if the cluster is matched to reconstructed tracks that are consistent with originating from a photon conversion and have a reconstructed conversion vertex.

Electron candidates are reconstructed from energy deposits in the electromagnetic calorimeter associated with reconstructed tracks from the ID system. They are identified with a combined likelihood technique and required to satisfy the \emph{MediumLH} identification criteria~\cite{EGAM-2018-01} and have a pseudorapidity of  $|\eta_{\text{cluster}}|<2.47$, excluding candidates in the transition region (${|\eta_{\text{cluster}}|\not\in[1.37,1.52]}$). Muon candidates are reconstructed by combining track segments in various layers of the muon spectrometer and tracks in the ID system. Muons are required to satisfy the \emph{Medium} identification quality criteria and to have $|\eta| < 2.5$~\cite{MUON-2018-03}. Both electron and muon candidates are required to have $\pT>$~7~\GeV and meet the
loose working point (WP) requirement of the prompt-lepton isolation discriminant~\cite{HIGG-2017-02}. The transverse impact parameter divided by its estimated uncertainty is required to be less than five (three) for electron (muon) candidates and the longitudinal impact
parameter must be smaller than 0.5~mm for both lepton flavours.

Jets are reconstructed based on particle-flow objects using the anti-$k_{t}$ algorithm~\cite{Cacciari:2008gp} based on the \textsc{FastJet} implementation~\cite{Fastjet} with a radius parameter $R=0.4$~\cite{PERF-2015-09}. The jet energy scale (JES) and resolution are calibrated using simulations with in situ corrections obtained from data~\cite{JETM-2018-05}. They are required to have $\pT > \SI{25}{\GeV}$ and $|\eta| < 2.5$. To reject jets from pile-up or other primary vertices, a \emph{jet vertex tagger} (JVT) discriminant~\cite{PERF-2014-03} is required to be larger than 0.59 for jets with $\pT < \SI{60}{\GeV}$ and $|\eta| < 2.4$. Jets arising from \bquark hadronisation, are identified using the DL1r \btag algorithm~\cite{FTAG-2019-07}. The $b$-tagged jets are required to satisfy a WP that corresponds to a 70\% per-jet efficiency for identifying $b$-quark-initiated jets in \ttbar simulated events.

The missing transverse momentum vector, with magnitude \met, corresponds to the negative sum of the transverse momenta of the reconstructed and calibrated physical objects, with a `soft term’ built from all other tracks associated with the primary vertex but not matched to a reconstructed object~\cite{JETM-2020-03}.

Events are selected if they have exactly two isolated photons and exactly one isolated electron or muon that must be matched to the corresponding trigger-level object. The \pt thresholds for the matched leptons are $25~\GeV$ for 2015 data, $27~\GeV$ for 2016 data, and $28~\GeV$ for 2017 and 2018 data. Events are rejected if additional lepton candidates with $\pt > 7~\GeV$ are present. Furthermore, events where the invariant mass of the electron and any photon is within the range 86.19--96.19~\GeV are removed, to suppress the background from $Z \to ee$ events, where an electron is misidentified as a photon. The last requirement for retaining an event is that it must contain at least four reconstructed jets, with at least one of them tagged as a $b$-jet.

After applying the full event selection, the background contribution accounts for roughly 50\% of the total events. Based on the photon origin, the background processes are divided into prompt photon background, where both photons are prompt, and fake photon background, where at least a photon is mimicked by another object, such as a misreconstructed hadron or electron. The contribution from events with misidentified leptons or leptons from heavy flavour decays is negligible and not further considered.

The prompt photon background is estimated by using MC simulations. The fraction of these background events is about 15\% of the total expected events. The largest prompt photon background contribution arises from $V$ boson events with two photons, and \ttH events. In the following, \tty events passing the overlap removal are referred to as `\tty (prompt $\gamma$)' and prompt photon background events from the single-top quark, single and diboson samples, $t\bar{t}V$, $t\bar{t}H$ and \ttbar simulated samples as the `\catprompty' background category. The largest contributions to this category arise from $V$ boson events with two photons (35\%), and \ttH events (22\%).

In this paper, the MC estimate of events with fake photons is corrected using the data-driven scale factors measured in dedicated control regions in Ref.~\cite{TOPQ-2018-24}. The contribution from processes with at least one electron mimicking a photon signature, `e-fake', originates mainly from dilepton \tty events where an electron from a top-quark decay is identified as a photon yielding the same signature as the single-lepton \ttyy channel. In particular, those events amount to 80\% of the total events in this category. The e-fake background is estimated in data from the fraction of electron--positron candidates from $Z\rightarrow e^+e^-$ decays that are reconstructed as $e\gamma$ pairs~\cite{EGAM-2021-01}. The ratio of the e-fake rates in data and simulation is used to correct the e-fake background predicted by the simulation in the signal region. The scale factors are applied per photon as a function of the fake photon \pt and $\eta$, separately for converted and unconverted photons. After applying the data-driven corrections, this background category represents about 20\% of the total number of selected events. The background contribution from events where at least one photon candidate arises from misidentified hadrons or from hadron decays, referred to as `h-fake', was estimated from data using the `ABCD' data-driven method~\cite{CDF:1990kbl} in Ref.~\cite{TOPQ-2018-24}. It is also measured as a
function of photon \pt, $\eta$ and conversion type. This category corresponds to about 10\% of the events, where 80\% of them correspond to \tty events with a fake photon. The contribution of events with one e-fake and one h-fake photon represents about 1\% of the events and is included in the e-fake contribution. Example variables illustrating the description of the data by the simulation after the event selection are shown in Section~\ref{sec:measurements}.

\section{Systematic uncertainties}
The measurements are affected by systematic uncertainties owing to theoretical assumptions and detector effects and uncertainties arising from the limited number of events in the MC simulations. The determination of the modelling and normalisation uncertainties associated with the background samples and the experimental uncertainties follows the same prescriptions as in Ref.~\cite{TOPQ-2018-24}.

The signal and background modelling uncertainties include effects due to the choice of renormalisation and factorisation scales, PDF set, parton shower, amount of QCD initial-state radiation (ISR) and final-state radiation (FSR). Modelling uncertainties are treated as uncorrelated sources of uncertainty between different processes because they are simulated using different generators or at different accuracy in QCD\@. The effect of the QCD scale uncertainty for the \ttyy, \tty production and decay, \Wty and \ttbar processes is evaluated independently by separately halving and doubling the renormalisation and factorisation scales relative to the nominal scale choice. The uncertainty in the PDFs for the \ttyy signal and the main background processes, \tty production and decay, are obtained using the 30 PDF variations of the PDF4LHC15 prescription~\cite{Butterworth:2015oua}. The uncertainty from the parton shower and hadronisation for  \ttyy, \tty production and decay, \Wty and \ttbar processes is estimated by comparing the nominal simulated samples interfaced with \PYTHIA[8] with alternative samples interfaced to \HERWIG[7.2.1]~\cite{Bahr:2008pv,Bellm:2015jjp}. Uncertainties due to the value of \alphas used in the ISR parton shower modelling are estimated by comparing the nominal \ttyy, \tty, and \ttbar simulations to alternative samples generated with varied radiation parameter settings in the A14 tune, controlled by the \textit{var3c} parameter implemented in \PYTHIA[8]~\cite{ATL-PHYS-PUB-2014-021}. This parameter is varied within its uncertainties corresponding to variation of $\alphas(m_Z)$ between 0.115 and 0.140. The impact of the FSR uncertainties in \ttyy is evaluated by varying the renormalisation scale for QCD emission in the FSR in \pythia8 by factors of 0.5 and 2.0, respectively.  For the \ttbar process, an additional ISR uncertainty is obtained by comparing the nominal sample with an additional one with the \hdamp parameter varied by a factor of two~\cite{ATL-PHYS-PUB-2018-009} and the impact of the ME corrections applied to all emissions is evaluated by comparing the nominal \ttbar sample to a dedicated \powheg + \PYTHIA[8] sample with the corrections turned off.

Additionally, normalisation uncertainties are considered for the different prompt-photon background processes. A normalisation uncertainty of 7\% is considered separately for the \tty production and \tty decay samples, which is based on the measured uncertainties for those processes in Ref.~\cite{TOPQ-2018-24}. A 6\% normalisation uncertainty is assigned to the \ttbar simulation~\cite{Czakon:2011xx}, which corresponds to the uncertainty of the NNLO in perturbative QCD, including soft-gluon resummation to next-to-next-to-leading-log. A 10\% uncertainty is assigned to the $t\bar{t}H$ process based on the calculations from Ref.~\cite{deFlorian:2016spz}. For $V\gamma\gamma$ and $V\gamma$ events a 50\% normalisation uncertainty is assigned. While the processes are measured with a 20\% precision or better~\cite{STDM-2018-33}, this normalisation uncertainty accounts for the possible mismodelling of the heavy flavour jets. The value is based on the differences between data and simulation observed in dedicated $W\gamma$ validation regions in Ref.~\cite{TOPQ-2021-29}. For the small background processes, single top quark (t- and s-channel), $t\bar{t}V$, $V\gamma$, $V+\mathrm{jets}$, $tW\gamma$ and diboson processes, a 50\% uncertainty is assigned, following the prescription of the latest \tty cross-section measurement~\cite{TOPQ-2018-24}.
For the backgrounds estimated with data-driven methods, i.e., e-fakes and h-fakes, the corresponding uncertainties of the scale factors applied to the simulation are considered~\cite{TOPQ-2018-24}. The sources of systematic uncertainty in the e-fake background estimate are related to the choice of the function describing the $Z$ boson mass peak, the mass range considered, and from the assumed function used to describe the background. The systematic uncertainties in the \cathfake background include the modelling of the \ttbar process, which accounts for most of the \cathfake events in the dedicated control regions in the \tty measurement, and the normalisation of the e-fake and prompt background processes.

The experimental systematic uncertainties account for uncertainties related to the reconstruction and identification of the physics objects, the integrated luminosity and the simulation of pile-up events. Photon and lepton identification and isolation efficiencies, momentum scale and resolution~\cite{EGAM-2018-01,EGAM-2021-01,MUON-2018-03}, and lepton trigger efficiencies in the simulation are corrected using scale factors to improve the description of the data. The associated systematic uncertainties are obtained by varying the corrections within their uncertainties.

The JES uncertainty is derived from a combination of simulations, test-beam data and in situ measurements~\cite{JETM-2018-05} resulting in 30 uncorrelated JES uncertainty subcomponents, which take into account effects from jet-flavour composition, $\eta$-intercalibration, punch-through, single-particle response, calorimeter response to different jet flavours, and pile-up. The jet energy resolution in simulation is smeared by its corresponding uncertainty~\cite{JETM-2018-05} split into 13 uncorrelated sources. The uncertainty associated with the JVT discriminant for pile-up jet rejection is obtained by varying the efficiency correction factors~\cite{PERF-2014-03}.

The uncertainties in the $b$-jet tagging calibration are determined separately for $b$-jets, $c$-jets and light-flavour jets~\cite{FTAG-2018-01,FTAG-2020-08,FTAG-2019-02}. For each jet category, the uncertainties are decomposed into several uncorrelated components. Additionally, a 50\% normalisation uncertainty is applied to events with high-\pt jets above the \pt limit where the $b$-jet tagging algorithm is calibrated.

The uncertainty in \MET results from the propagation of the uncertainties in the energy scales and resolutions of photons, leptons and jets, and the modelling of its soft term~\cite{JETM-2020-03}.

The uncertainty in the total integrated luminosity, using the LUCID-2 detector \cite{LUCID2} for the primary luminosity measurements, is estimated to be 0.83\%~\cite{DAPR-2021-01}. The uncertainty arising from the modelling of pile-up is determined by varying the pile-up reweighting applied to correct the simulation to better describe the data within its uncertainties.

\section{Measurement of the \ttyy cross section and \ttyy/\tty cross-section ratio}
\label{sec:measurements}
A multivariate analysis using a BDT is performed to separate the \ttyy signal from the background processes. The training is performed with the eXtreme Gradient Boosting classifier from the XGBoost library~\cite{10.1145/2939672.2939785}. A 4-fold cross-validation was performed, where the events were randomly split into four equally sized subsets. For each fold, three subsets were used for training the BDT configuration, and the remaining one was used for testing. %
The event weights are applied to the events in the training and testing samples. Events with negative weights in samples generated at NLO are included in the training using the absolute value of their weights. It was verified that the shape of the input variables is not significantly affected. The training uses simulated events from the \ttyy, \tty production and decay and $V\gamma\gamma$ samples, with the corresponding data-driven scale factors applied to the fake photon events. The \ttyy signal and the total background contributions are rescaled to be equally represented in the training. A total of 19 variables are selected for the training based on their good agreement between data and MC simulation, balanced correlations among them, yielding good separation and best expected significance. They are related to the properties of the photons, such as photon conversion type, \pt and $\eta$, invariant masses and angular separations of different objects in the event (e.g.\ to the invariant mass and $\Delta R$ of the two photons, the lepton and each photon and the photons and the closest $b$-tagged jet), to the \met and to the jet properties, such as the number of $b$-tagged jets. It was checked that the simulation accurately describes the data within the uncertainties for all input variables. Example variables with the most separation power are shown in Figure~\ref{fig:prefit_variables_1} before the fit.

\begin{figure}
\centering
\subfloat[]{\includegraphics[width=0.33\textwidth]{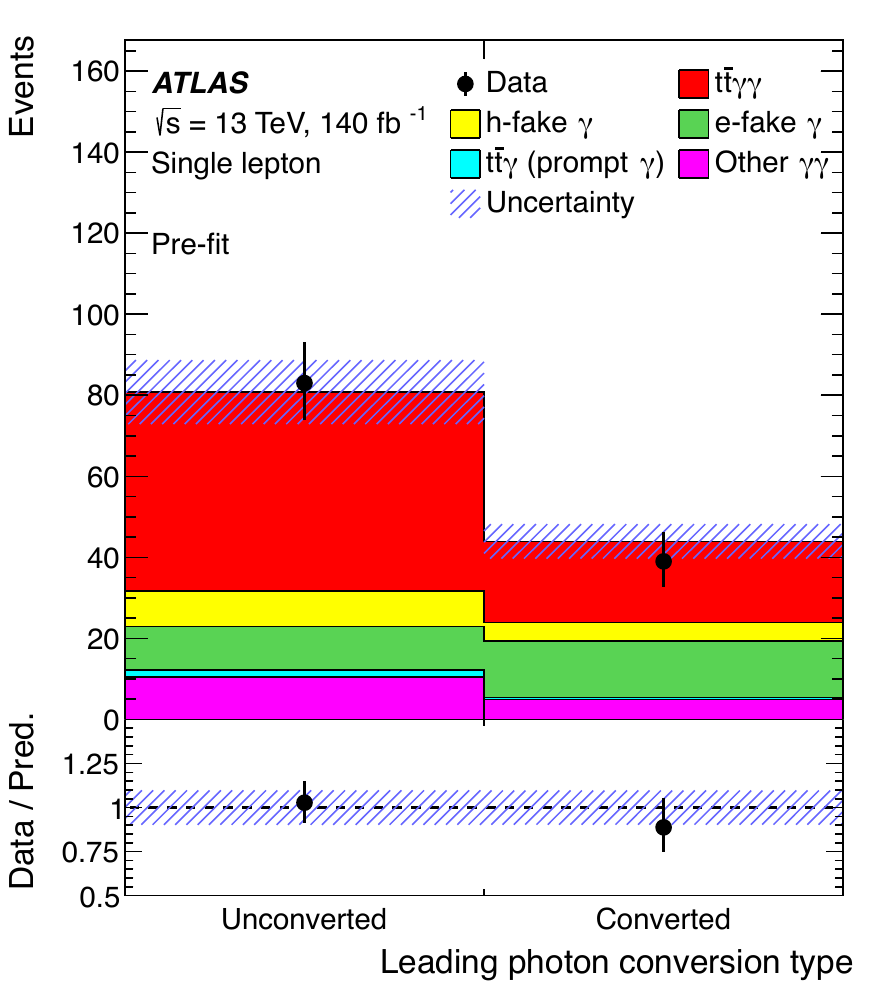}}
\subfloat[]{\includegraphics[width=0.33\textwidth]{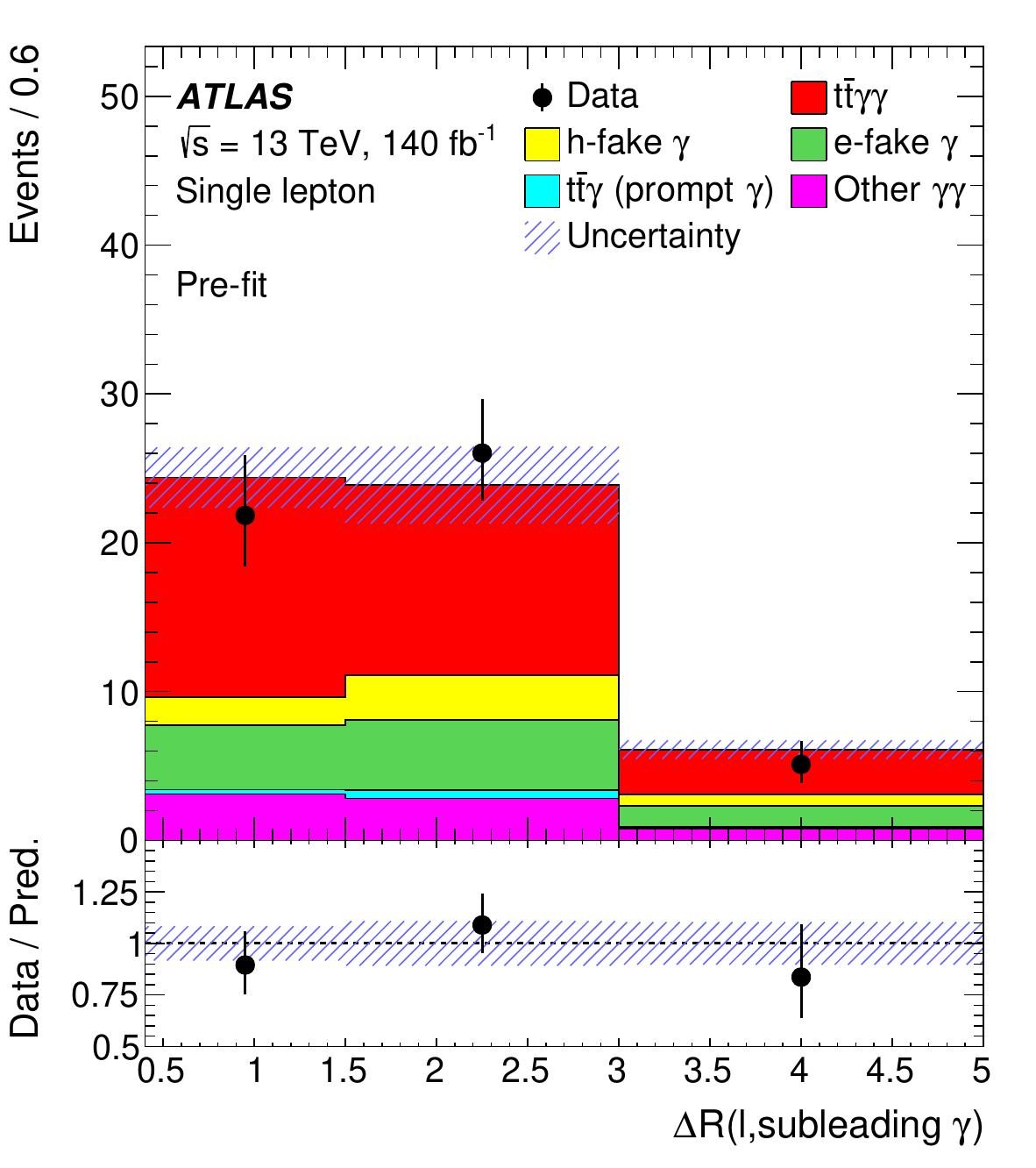}}
\subfloat[]{\includegraphics[width=0.33\textwidth]{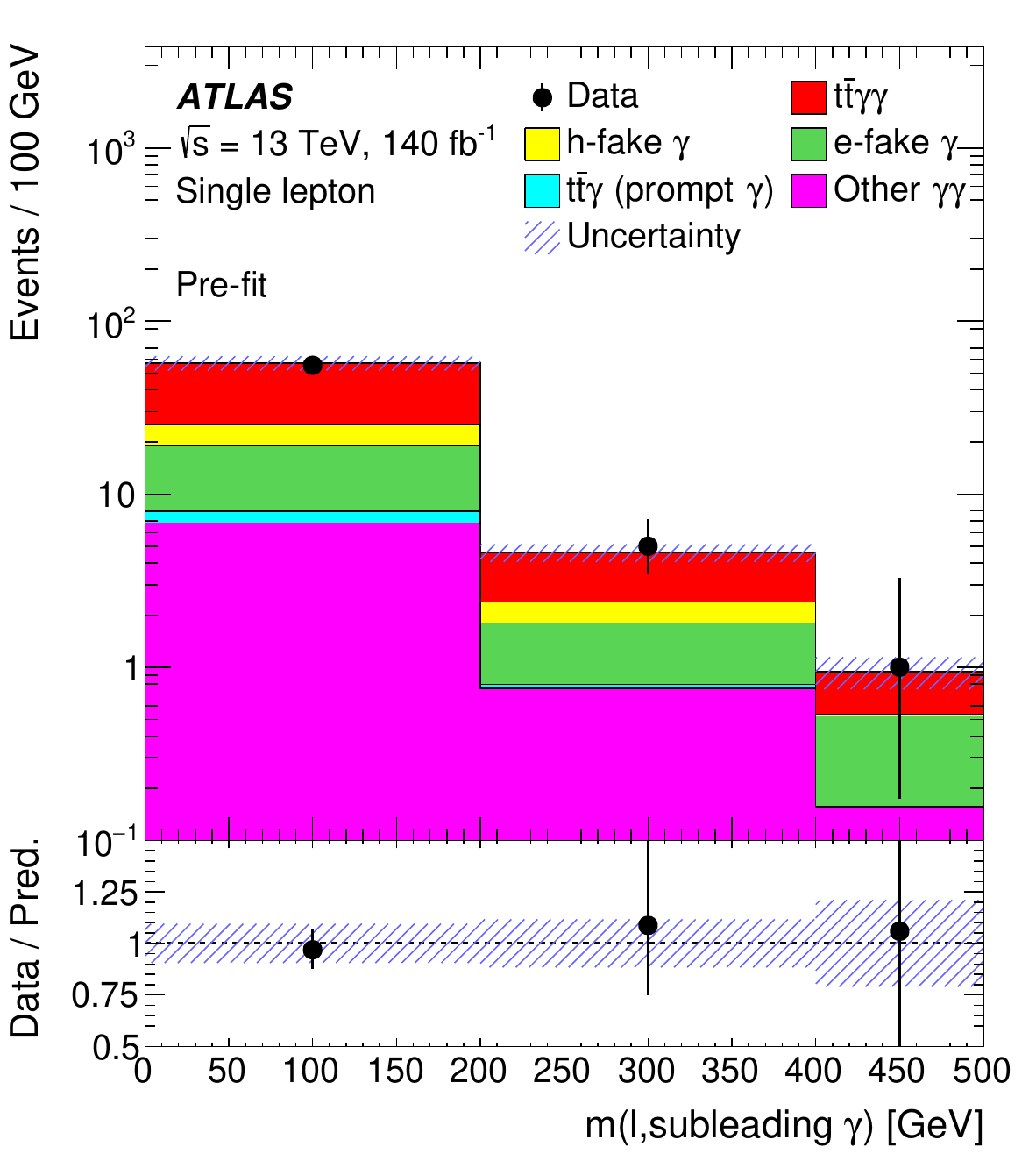}}
\caption{Example of input variables with large separation power, before the fit to data: (a) conversion type of the photon leading in \pt, (b) $\Delta R$ between the lepton and the subleading photon in \pt and (c) their invariant mass. The uncertainty band corresponds to the total uncertainties before the fit to data. The lower part of the plot shows the ratio of the data to the prediction. If present, overflow events are included in the last bin of the distribution. The `\catprompty' category includes simulated events with two prompt photons from single-top quark, single and diboson processes, \(t\bar{t}V\), \(t\bar{t}H\), and \(\ttbar\) samples. The normalisation of the \ttyy sample is rescaled using the NLO/LO $K$-factor described in Section~\ref{sec:simulations}.}
\label{fig:prefit_variables_1}
\end{figure}

The measurements are performed at particle level in a fiducial phase space, which is defined based on the requirements used in Ref.~\cite{TOPQ-2018-24} and closely follows the kinematic requirements at reconstruction level. Photons are required to not originate from a hadron decay, to have $p_T > \SI{20}{\GeV}$ and $|\eta| < \SI{2.37}{}$ and to be isolated such that the sum of transverse momenta of all charged particles surrounding the photon within $\Delta R \leq$ 0.2 must be smaller than 5\% of its \pt. Electrons and muons are dressed with nearby photons, which do not originate from hadrons, in a $\Delta R <0.1$ cone around the lepton. Only leptons not originating from hadron decays are selected. Jets are clustered with the anti-$k_{t}$ algorithm with a radius of R = 0.4. All stable particles are considered in the clustering, except for the selected electrons, muons and photons, and the neutrinos originating from the top quarks. A particle-level jet is identified as a $b$-jet if a hadron with $\pT > \SI{5}{\GeV}$ containing a $b$-quark is matched to the jet through a ghost-matching method~\cite{Cacciari:2008gn}. Lepton and jets are required to have \pt $>$ 25 \GeV and $|\eta|<$ 2.5. Additionally, leptons within $\Delta R < 0.4$ from a jet are excluded and jets are removed if they are within $\Delta R = 0.4$ of an isolated photon candidate. Events are additionally required to satisfy that the $\Delta R$ distance between the lepton and any photon and between the photons is larger than 0.4. The fiducial phase space is defined by requiring exactly two photons, exactly one electron or muon, and at least four jets among which at least one is a $b$-jet. Events are rejected if there are additional leptons with $\pt > 7$~\GeV. The fraction of simulated \ttyy signal events that pass the event selection at reconstruction level over the number of events in the fiducial region at particle level is 19\% and the fraction of signal events that fulfil the selection at reconstruction level and the particle-level requirements is 79\%.

The fiducial cross section at particle level is measured using a profile likelihood fit to the BDT output that profiles nuisance parameters and minimises the negative log-likelihood~\cite{Cowan:2010js}. The normalisation of the \ttyy signal template is treated as the parameter of interest in the fit. The resulting normalisation factor is then used to extract the observed cross section by scaling the predicted cross section in the fiducial phase space at particle level. Background processes are modelled using MC simulation and data-driven methods, with normalisation and shape uncertainties included as nuisance parameters in the likelihood fit. All MC samples used to evaluate signal modelling uncertainties are scaled to the same cross section in the fiducial phase space. Each source of systematic uncertainty, including those affecting the signal, is incorporated as a Gaussian prior in the likelihood function. The variations are symmetrised, taking half the difference between the upward and downward variations as the uncertainty. If both variations have the same sign, the average of the difference between each variation and the nominal value is chosen as the two-sided uncertainty. If only one variation is available (e.g. the uncertainty from the parton shower or the PDF uncertainties) the difference between the nominal and alternative prediction is taken as the size of the uncertainty.
A pruning procedure is implemented to remove uncertainties that are too small to meaningfully impact the results, improving the fit's convergence and reducing computing time. %
In addition, a smoothing procedure is applied, where necessary, to reduce the impact of statistical fluctuations in the systematic templates. The impact of these procedures on the precision of the results is negligible. The output of the BDT after the fit is shown in Figure~\ref{fig:postfit_ttyy}. The binning is optimised to enhance signal-to-background separation while ensuring smooth templates for the small background contributions. The fraction of background events after the fit corresponds to 47\%. Good agreement is observed between the data and the prediction. The global goodness of fit, evaluated with a \textit{saturated model}~\cite{BAKER1984437}, yields a $p$-value of 49\%.

\begin{figure}
\centering
\includegraphics[width=0.45\textwidth]{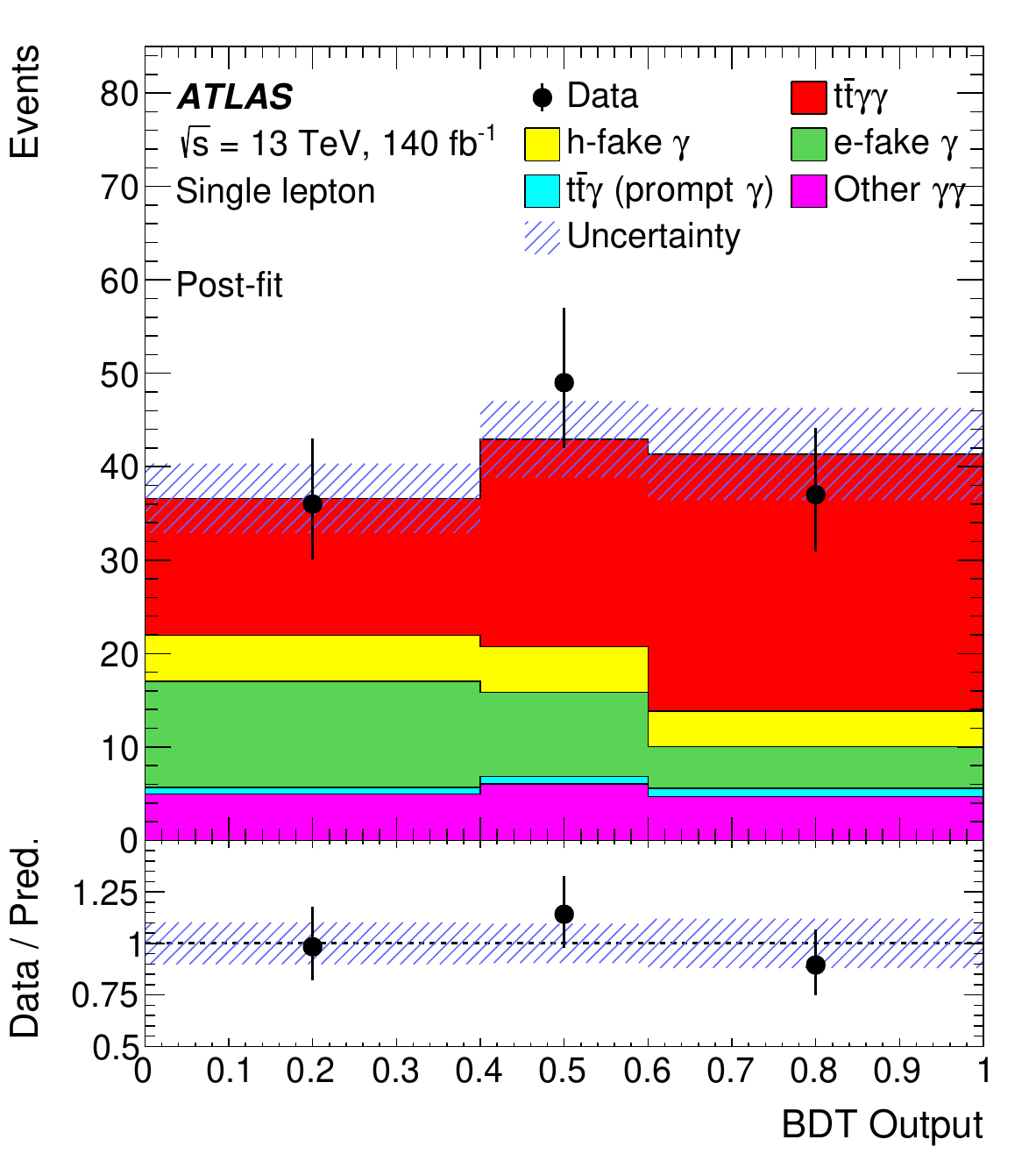}
\caption{BDT distribution output after the fit to data. The uncertainty band corresponds to the total uncertainties with correlations among uncertainties taken into account as determined in the fit. The lower
part of the plot shows the ratio of the data to the prediction. The `\catprompty' category includes simulated events with two prompt photons from single-top quark, single and diboson processes, \(t\bar{t}V\), \(t\bar{t}H\), and \(\ttbar\) samples.
}
\label{fig:postfit_ttyy}
\end{figure}

The fiducial cross section at particle level %
is found to be
\begin{equation}
\nonumber
\sigma_{t\bar{t}\gamma\gamma} = 2.42^{+0.58}_{-0.53}\, \text{fb} = 2.42^{+0.46}_{-0.38} \, ({ \text{stat}})\, ^{+0.35}_{-0.38} ({\mathrm{syst}})\, \text{fb}.
\end{equation}
The total relative uncertainty of the measurement is approximately 23\%, with the dominant contribution coming from the statistical uncertainty of the data, which is about 17\%. Systematic uncertainties contribute approximately 15\% to the total relative uncertainty. The systematic uncertainties are dominated by the experimental uncertainties related to the jets and the photons, and the uncertainties in the fake-photon background estimates, followed by the normalisation uncertainties of the other $\gamma\gamma$ background contribution. The impact of the different categories of systematic uncertainties is summarised in Table~\ref{tab:grouped_uncertainties} (left column). None of the nuisance parameters are significantly constrained by the fit. Their best-fit values are well within one standard deviation of their initial estimates. The prediction for the fiducial \ttyy cross section given by the nominal \ttyy \MGNLO+\Pythia8 LO simulation is $1.53^{+0.40}_{-0.52} ({ \text{scale}})^{+0.05}_{-0.03} ({\text{PDF}})\, \text{fb}$. The observed significance with respect to the background-only hypothesis is 5.2 standard deviations. The expected significance evaluated with the LO simulation is 3.8 standard deviations, assessed with toy experiments. The expected significance is 5.7 standard deviations with the approximate $K$-factor of 1.7, described in Section~\ref{sec:simulations}, which is used for the prior normalisation of the \ttyy sample in the distributions before the fit to data. An additional check was performed by fitting the number of events instead of the BDT output. The result, $\sigma_{t\bar{t}\gamma\gamma} = 2.51^{+0.62}_{-0.58}\, \text{fb}$, is consistent with the one from the default fit.

The ratio of the cross sections of the \ttyy process and the \tty process (\tty production and decay) in the single-lepton channel is additionally extracted at particle level. The input distributions used to measure the \tty cross sections are taken from Ref.~\cite{TOPQ-2018-24}. The event reconstruction and selection are designed to avoid any statistical overlap with the \ttyy measurement. The selection criteria at reconstruction level and particle-level phase space definition are analogous to those of the \ttyy analysis, except that exactly one isolated photon is required instead of two. Neural networks (NNs) are used to enhance the separation of $t\bar{t}\gamma$ production and the various background processes. A four-class NN is used to define a signal region enriched in $t\bar{t}\gamma$ production events and three control regions enriched in $t\bar{t}\gamma$ decay, photon fakes and other events with prompt photons, respectively. The same object selection and the same sources of systematic uncertainties are considered in both measurements. Therefore, the uncertainties related to the object reconstruction, luminosity and pile-up modelling are kept fully correlated between the measurements. The uncertainties related to the modelling and the normalisation of the common processes,  which are estimated with the same set of MC samples (\tty production, \tty decay, \ttbar, \tWy, t-channel and s-channel single top quark, \ttV, $V\gamma$, $V$+jets, inclusive diboson samples) are also correlated among the two measurements. Common background normalisation uncertainties and the uncertainties in the data-driven \efake and \hfakes are also correlated. Modelling uncertainties between different processes (e.g. PS uncertainties between \tty production and decay and \ttyy) are considered uncorrelated as in the individual analyses.

The ratio of \ttyy to \tty cross section is obtained by fitting simultaneously the \ttyy BDT output and the variables used in the four regions of the \tty measurement. In particular, the output of the corresponding NN classifiers are the input distributions in the \tty production and the \tty decay regions, while the distributions of the second largest pseudo-continuous $b$-tagging score are used in the regions enriched in photon fakes and in other prompt photons~\cite{TOPQ-2018-24}. In the measurement, the cross-section ratio is treated as the parameter of interest, while the \tty cross section is profiled using data. The \tty production and \tty decay templates are added together, and the normalisation of the combined template is free floating in the fit. A 20\% normalisation uncertainty is considered for the \tty decay template, which is the pre-fit uncertainty associated with the normalisation of \tty decay in the total \tty measurement~\cite{TOPQ-2018-24}. This value is based on the estimated uncertainty in the NLO $K$-factor used to correct the normalisation of \tty decay LO sample, as obtained in Ref.~\cite{TOPQ-2021-29}. This nuisance parameter allows the fit to adjust the fraction of \tty production and \tty decay events.

\begin{table}
\centering
\caption{Summary of the relative impact of all the systematic uncertainties, in percentage, on the \ttyy fiducial
inclusive cross section and $R_{t\bar{t}\gamma\gamma/t\bar{t}\gamma}$ grouped into
different categories. The category ‘Jet’ corresponds to the effect of JES, jet resolution
and JVT uncertainties, ‘Photon’ and ‘Leptons’ include all experimental
uncertainties related to photons and leptons (including trigger uncertainties), respectively.}
\label{tab:grouped_uncertainties}
\begin{tabular}{lcc}
\toprule
Source & {\footnotesize $\Delta\sigma_{t\bar{t}\gamma\gamma}/\sigma_{t\bar{t}\gamma\gamma}$ [\%]} &  {\footnotesize $\Delta R_{t\bar{t}\gamma\gamma/t\bar{t}\gamma}/R_{t\bar{t}\gamma\gamma/t\bar{t}\gamma}$ [\%]} \\
\midrule
\ttyy modelling & 1.4 & 1.3 \\
Prompt-photon background norm. \& modelling & 4.4 & 5.8 \\
Fake-photon background estimates & 6.5 & 0.5 \\
Fake-lepton background estimate & -- & 0.9 \\
\midrule
Jet & 9.7 & 5.9 \\
Photon & 6.5 & 4.0 \\
$b$-tagging & 3.4 & 1.0 \\
Leptons & 1.5 & 0.3 \\
Luminosity & 1.4 & 0.1 \\
\met & 0.4 & 1.1 \\
Pile-up & 1.6 & 1.4 \\
\midrule
MC statistical uncertainties & 2.5 & 2.8 \\
\midrule
Total systematic uncertainty & 15.0\phantom{0} & 10.0\phantom{0} \\
\bottomrule
\end{tabular}

\end{table}

The measurement yields $R_{t\bar{t}\gamma\gamma/t\bar{t}\gamma} = (3.30^{+0.70}_{-0.65}) \times 10^{-3} = (3.30^{+0.63}_{-0.55}\textrm{ (stat)}^{+0.32}_{-0.34}\textrm{ (syst)}) \times 10^{-3}$. The precision of the measurement of about 20\% is dominated by the statistical uncertainty of the data, which is approximately 18\%, while the total systematic uncertainty is about 10\%. The impact of correlated systematic uncertainties partially cancels in the ratio. Specifically, systematic uncertainties that cancel significantly include experimental sources associated with jets, $b$-tagging, photons, and luminosity, and those arising from data-driven estimates of fake-photon backgrounds. As expected, the relative impact of \tty production and decay modelling uncertainties increases compared with the measurement of the \ttyy cross section, as shown in Table~\ref{tab:grouped_uncertainties} (right column). The measured value and total uncertainty of the \tty cross section agree with the result from Ref.~\cite{TOPQ-2018-24}.

There are no available calculations of the ratio that would allow for a direct comparison with the measured value. As a reference, using the nominal \MGNLO+\PYTHIA[8] samples, the ratio of the fiducial \ttyy LO cross section to the \tty cross section---defined as the sum of the NLO \tty production and \tty decay contributions, including the NLO/LO \(K\)-factor---yields $(2.46 ^{+0.73}_{-0.94}\,({ \text{scale}})^{+0.10}_{-0.06} \,({\text{PDF}})) \times 10^{-3}$. The scale and PDF uncertainties are evaluated assuming uncorrelated uncertainties between the individual cross sections. This value and its uncertainty should be treated as illustrative, as they are based on a combination of LO and NLO predictions, and no uncertainty is assigned to the $K$-factor applied to the \tty decay contribution.

\section{Conclusion}
This paper presents the observation of the production of top quark pairs in association with two photons. The measurement is performed using a data sample of proton--proton collisions collected with the ATLAS detector at the LHC at $\sqrt{s}$ = 13 \TeV, corresponding to a total integrated luminosity of \intlumitot. The cross-section measured in a fiducial phase space defined at particle level, with kinematic requirements on the photons, lepton, and jets, yields $2.42^{+0.58}_{-0.53}\,\text{fb}$, with an observed significance of 5.2 standard deviations. The available MC prediction at LO underestimate the cross section as expected. In the measurement of the ratio of the \ttyy and the \tty processes, several systematic uncertainties—particularly those related to jets, $b$-tagging, photons, and luminosity—partially cancel out. The ratio, measured in the fiducial phase space, is found to be $(3.30^{+0.70}_{-0.65})\times 10^{-3}$.

\section*{Acknowledgements}
%

%

%
%

%
%

We thank CERN for the very successful operation of the LHC and its injectors, as well as the support staff at
CERN and at our institutions worldwide without whom ATLAS could not be operated efficiently.

The crucial computing support from all WLCG partners is acknowledged gratefully, in particular from CERN, the ATLAS Tier-1 facilities at TRIUMF/SFU (Canada), NDGF (Denmark, Norway, Sweden), CC-IN2P3 (France), KIT/GridKA (Germany), INFN-CNAF (Italy), NL-T1 (Netherlands), PIC (Spain), RAL (UK) and BNL (USA), the Tier-2 facilities worldwide and large non-WLCG resource providers. Major contributors of computing resources are listed in Ref.~\cite{ATL-SOFT-PUB-2025-001}.

We gratefully acknowledge the support of ANPCyT, Argentina; YerPhI, Armenia; ARC, Australia; BMWFW and FWF, Austria; ANAS, Azerbaijan; CNPq and FAPESP, Brazil; NSERC, NRC and CFI, Canada; CERN; ANID, Chile; CAS, MOST and NSFC, China; Minciencias, Colombia; MEYS CR, Czech Republic; DNRF and DNSRC, Denmark; IN2P3-CNRS and CEA-DRF/IRFU, France; SRNSFG, Georgia; BMFTR, HGF and MPG, Germany; GSRI, Greece; RGC and Hong Kong SAR, China; ICHEP and Academy of Sciences and Humanities, Israel; INFN, Italy; MEXT and JSPS, Japan; CNRST, Morocco; NWO, Netherlands; RCN, Norway; MNiSW, Poland; FCT, Portugal; MNE/IFA, Romania; MSTDI, Serbia; MSSR, Slovakia; ARIS and MVZI, Slovenia; DSI/NRF, South Africa; MICIU/AEI, Spain; SRC and Wallenberg Foundation, Sweden; SERI, SNSF and Cantons of Bern and Geneva, Switzerland; NSTC, Taipei; TENMAK, T\"urkiye; STFC/UKRI, United Kingdom; DOE and NSF, United States of America.

Individual groups and members have received support from BCKDF, CANARIE, CRC and DRAC, Canada; CERN-CZ, FORTE and PRIMUS, Czech Republic; COST, ERC, ERDF, Horizon 2020, ICSC-NextGenerationEU and Marie Sk{\l}odowska-Curie Actions, European Union; Investissements d'Avenir Labex, Investissements d'Avenir Idex and ANR, France; DFG and AvH Foundation, Germany; Herakleitos, Thales and Aristeia programmes co-financed by EU-ESF and the Greek NSRF, Greece; BSF-NSF and MINERVA, Israel; NCN and NAWA, Poland; La Caixa Banking Foundation, CERCA Programme Generalitat de Catalunya and PROMETEO and GenT Programmes Generalitat Valenciana, Spain; G\"{o}ran Gustafssons Stiftelse, Sweden; The Royal Society and Leverhulme Trust, United Kingdom.

In addition, individual members wish to acknowledge support from CERN: European Organization for Nuclear Research (CERN DOCT); Chile: Agencia Nacional de Investigaci\'on y Desarrollo (FONDECYT 1230812, FONDECYT 1240864, Fondecyt 3240661, Fondecyt Regular 1240721); China: Chinese Ministry of Science and Technology (MOST-2023YFA1605700, MOST-2023YFA1609300), National Natural Science Foundation of China (NSFC - 12175119, NSFC 12275265); Czech Republic: Czech Science Foundation (GACR - 24-11373S), Ministry of Education Youth and Sports (ERC-CZ-LL2327, FORTE CZ.02.01.01/00/22\_008/0004632), PRIMUS Research Programme (PRIMUS/21/SCI/017); EU: H2020 European Research Council (ERC - 101002463); European Union: European Research Council (BARD No. 101116429, ERC - 948254, ERC 101089007), European Regional Development Fund (SMASH COFUND 101081355, SLO ERDF), Horizon 2020 Framework Programme (MUCCA - CHIST-ERA-19-XAI-00), European Union, Future Artificial Intelligence Research (FAIR-NextGenerationEU PE00000013), Italian Center for High Performance Computing, Big Data and Quantum Computing (ICSC, NextGenerationEU); France: Agence Nationale de la Recherche (ANR-21-CE31-0022, ANR-22-EDIR-0002, ANR-24-CE31-0504-01); Germany: Baden-Württemberg Stiftung (BW Stiftung-Postdoc Eliteprogramme), Deutsche Forschungsgemeinschaft (DFG - 469666862, DFG - CR 312/5-2); China: Research Grants Council (GRF); Italy: Istituto Nazionale di Fisica Nucleare (ICSC, NextGenerationEU), Ministero dell'Università e della Ricerca (NextGenEU 153D23001490006 M4C2.1.1, NextGenEU I53D23000820006 M4C2.1.1, NextGenEU I53D23001490006 M4C2.1.1, SOE2024\_0000023); Japan: Japan Society for the Promotion of Science (JSPS KAKENHI JP22H01227, JSPS KAKENHI JP22H04944, JSPS KAKENHI JP22KK0227, JSPS KAKENHI JP24K23939, JSPS KAKENHI JP25H00650, JSPS KAKENHI JP25H01291, JSPS KAKENHI JP25K01023); Norway: Research Council of Norway (RCN-314472); Poland: Ministry of Science and Higher Education (IDUB AGH, POB8, D4 no 9722), Polish National Science Centre (NCN 2021/42/E/ST2/00350, NCN OPUS 2023/51/B/ST2/02507, NCN OPUS nr 2022/47/B/ST2/03059, NCN UMO-2019/34/E/ST2/00393, UMO-2022/47/O/ST2/00148, UMO-2023/49/B/ST2/04085, UMO-2023/51/B/ST2/00920, UMO-2024/53/N/ST2/00869); Portugal: Foundation for Science and Technology (FCT); Spain: Ministry of Science and Innovation (MCIN \& NextGenEU PCI2022-135018-2, MICIN \& FEDER PID2021-125273NB, RYC2019-028510-I, RYC2020-030254-I, RYC2021-031273-I, RYC2022-038164-I); Sweden: Carl Trygger Foundation (Carl Trygger Foundation CTS 22:2312), Swedish Research Council (Swedish Research Council 2023-04654, VR 2021-03651, VR 2022-03845, VR 2022-04683, VR 2023-03403, VR 2024-05451), Knut and Alice Wallenberg Foundation (KAW 2018.0458, KAW 2022.0358, KAW 2023.0366); Switzerland: Swiss National Science Foundation (SNSF - PCEFP2\_194658); United Kingdom: Leverhulme Trust (Leverhulme Trust RPG-2020-004), Royal Society (NIF-R1-231091); United States of America: U.S. Department of Energy (ECA DE-AC02-76SF00515), Neubauer Family Foundation.

%
%


%
%
%
\printbibliography
\clearpage
 
\begin{flushleft}
\hypersetup{urlcolor=black}
{\Large The ATLAS Collaboration}

\bigskip

\AtlasOrcid[0000-0002-6665-4934]{G.~Aad}$^\textrm{\scriptsize 104}$,
\AtlasOrcid[0000-0001-7616-1554]{E.~Aakvaag}$^\textrm{\scriptsize 17}$,
\AtlasOrcid[0000-0002-5888-2734]{B.~Abbott}$^\textrm{\scriptsize 123}$,
\AtlasOrcid[0000-0002-0287-5869]{S.~Abdelhameed}$^\textrm{\scriptsize 119a}$,
\AtlasOrcid[0000-0002-1002-1652]{K.~Abeling}$^\textrm{\scriptsize 55}$,
\AtlasOrcid[0000-0001-5763-2760]{N.J.~Abicht}$^\textrm{\scriptsize 49}$,
\AtlasOrcid[0000-0002-8496-9294]{S.H.~Abidi}$^\textrm{\scriptsize 30}$,
\AtlasOrcid[0009-0003-6578-220X]{M.~Aboelela}$^\textrm{\scriptsize 45}$,
\AtlasOrcid[0000-0002-9987-2292]{A.~Aboulhorma}$^\textrm{\scriptsize 36e}$,
\AtlasOrcid[0000-0001-5329-6640]{H.~Abramowicz}$^\textrm{\scriptsize 157}$
% ...
\bigskip
\\

% ...
$^{17}$Department for Physics and Technology, University of Bergen, Bergen; Norway.\\
% ...
$^{30}$Physics Department, Brookhaven National Laboratory, Upton NY; United States of America.\\
% ...
$^{36}$$^{(a)}$Facult\'e des Sciences Ain Chock, Universit\'e Hassan II de Casablanca;$^{(b)}$Facult\'{e} des Sciences, Universit\'{e} Ibn-Tofail, K\'{e}nitra;$^{(c)}$Facult\'e des Sciences Semlalia, Universit\'e Cadi Ayyad, LPHEA-Marrakech;$^{(d)}$LPMR, Facult\'e des Sciences, Universit\'e Mohamed Premier, Oujda;$^{(e)}$Facult\'e des sciences, Universit\'e Mohammed V, Rabat;$^{(f)}$Institute of Applied Physics, Mohammed VI Polytechnic University, Ben Guerir; Morocco.\\
% ...
$^{45}$Physics Department, Southern Methodist University, Dallas TX; United States of America.\\
% ...
$^{49}$Fakult\"{a}t Physik , Technische Universit{\"a}t Dortmund, Dortmund; Germany.\\
% ...
$^{55}$II. Physikalisches Institut, Georg-August-Universit\"{a}t G\"ottingen, G\"ottingen; Germany.\\
% ...
$^{104}$CPPM, Aix-Marseille Universit\'e, CNRS/IN2P3, Marseille; France.\\
% ...
$^{119}$$^{(a)}$New York University Abu Dhabi, Abu Dhabi;$^{(b)}$United Arab Emirates University, Al Ain; United Arab Emirates.\\
% ...
$^{123}$Homer L. Dodge Department of Physics and Astronomy, University of Oklahoma, Norman OK; United States of America.\\
% ...
$^{157}$Raymond and Beverly Sackler School of Physics and Astronomy, Tel Aviv University, Tel Aviv; Israel.\\
% ...
$^{a}$ Also at Affiliated with an institute formerly covered by a cooperation agreement with CERN.\\
$^{b}$ Also at An-Najah National University, Nablus; Palestine.\\
$^{c}$ Also at Borough of Manhattan Community College, City University of New York, New York NY; United States of America.\\
$^{d}$ Also at Center for Interdisciplinary Research and Innovation (CIRI-AUTH), Thessaloniki; Greece.\\
$^{e}$ Also at Centre of Physics of the Universities of Minho and Porto (CF-UM-UP); Portugal.\\
$^{f}$ Also at CERN, Geneva; Switzerland.\\
% ...

\end{flushleft}


%

%
%

%

%
%
%
%

\end{document}